\documentclass[a4paper]{article}

\usepackage{icrc2013}

\title{Constraints on axion-like particles with H.E.S.S. from observations of PKS 2155-304}

\shorttitle{Constraints on axion-like particles with H.E.S.S. from observations of PKS 2155-304}

\authors{
Pierre Brun$^{1}$,
Denis Wouters$^{1}$,
for the H.E.S.S. Collaboration
}

\afiliations{
$^1$ CEA/Irfu, Centre de Saclay, F-91191 Gif-sur-Yvette, France
}

\email{pierre.brun@cea.fr}

\abstract{Axion-like particles are hypothetical new light (sub-eV) bosons predicted in some extensions of the Standard Model of particle physics. In astrophysical environments comprising high-energy gamma-rays and  turbulent magnetic fields, the existence of axion-like particles can modify the energy spectrum of the gamma-rays. This modification would take the form of an irregular behavior of the energy spectrum in a limited energy range. Data from the H.E.S.S. observations of the distant BL Lac PKS 2155-304 are used to derive conservative upper limits on the strength of the axion-like particle coupling to photons. This study gives rise to the first exclusions on axion-like particles from gamma-ray astronomy. The derived constraints apply to both light pseudo-scalar and scalar bosons that couple to the electromagnetic field.}


\begin{document}
\maketitle

\section{Introduction}

Pseudoscalar particles are predicted in many extension of the Standard Model. A well-known example is the axion, which is the pseudo Nambu-Goldstone boson associated to the breaking of the U(1) symmetry introduced by Peccei and Quinn to solve the strong CP problem~\cite{1977PhRvL..38.1440P, 1978PhRvL..40..279W, 1978PhRvL..40..223W}. Strong constraints on the energy scale of the breaking of the symmetry are obtained by astrophysical observations~\cite{2008LNP...741...51R}, implying that the mass of the axion should be lower than 1~eV. A generic property of pseudoscalar particles is their coupling to the electromagnetic field via a two-photon vertex. In the specific case of the axion, the coupling to photons is predicted to scale with its mass. However, more general low mass pseudoscalars with coupling and mass unrelated are expected from other exotic models~\cite{1987PhR...150....1K, 2012PDU.....1..116R, 2010ARNPS..60..405J}. Such particles are called axion-like particles (ALPs) and couple electromagnetically in the same way axions do.

The coupling of ALPs to two photons enables oscillations between photons and ALPs in an external magnetic field~\cite{1983PhRvL..51.1415S}. In $\gamma$-ray astronomy, these oscillations are sometimes considered as a possible mechanism responsible for the potential reduction of the opacity of the universe to TeV photons~\cite{2003JCAP...05..005C, 2012PhRvD..86g5024H, 2013PhRvD..87c5027M}. In~\cite{2012PhRvD..86d3005W}, a new, alternative ALP signature in the energy spectrum of bright high-energy photon sources has been proposed. Because of the turbulent nature of the astrophysical magnetic fields crossed by the photon beam of high-energy emitter, the observed spectrum can be affected by strong irregularities in a limited energy range. These irregularities are expected to appear around the critical energy $E_c = m^2/(2g_{\gamma a} B)$ above which the strong mixing regime is attained, where $m$ is the ALP mass, $g_{\gamma a}$ the coupling constant and $B$ the magnetic field strength. This method has been used in~\cite{2013arXiv1304.0989W} in the case of X-ray emission from the central source of the Hydra A galaxy cluster. This article reports the search for such irregularities in the energy spectrum of aTeV emitting source, which lies at the center of a cluster of galaxies. The dataset that is used has been obtained through observations with H.E.S.S. array of four ground-based Cherenkov telescopes, located in Namibia.

\begin{figure}
  \centering\includegraphics[width=1.05\columnwidth]{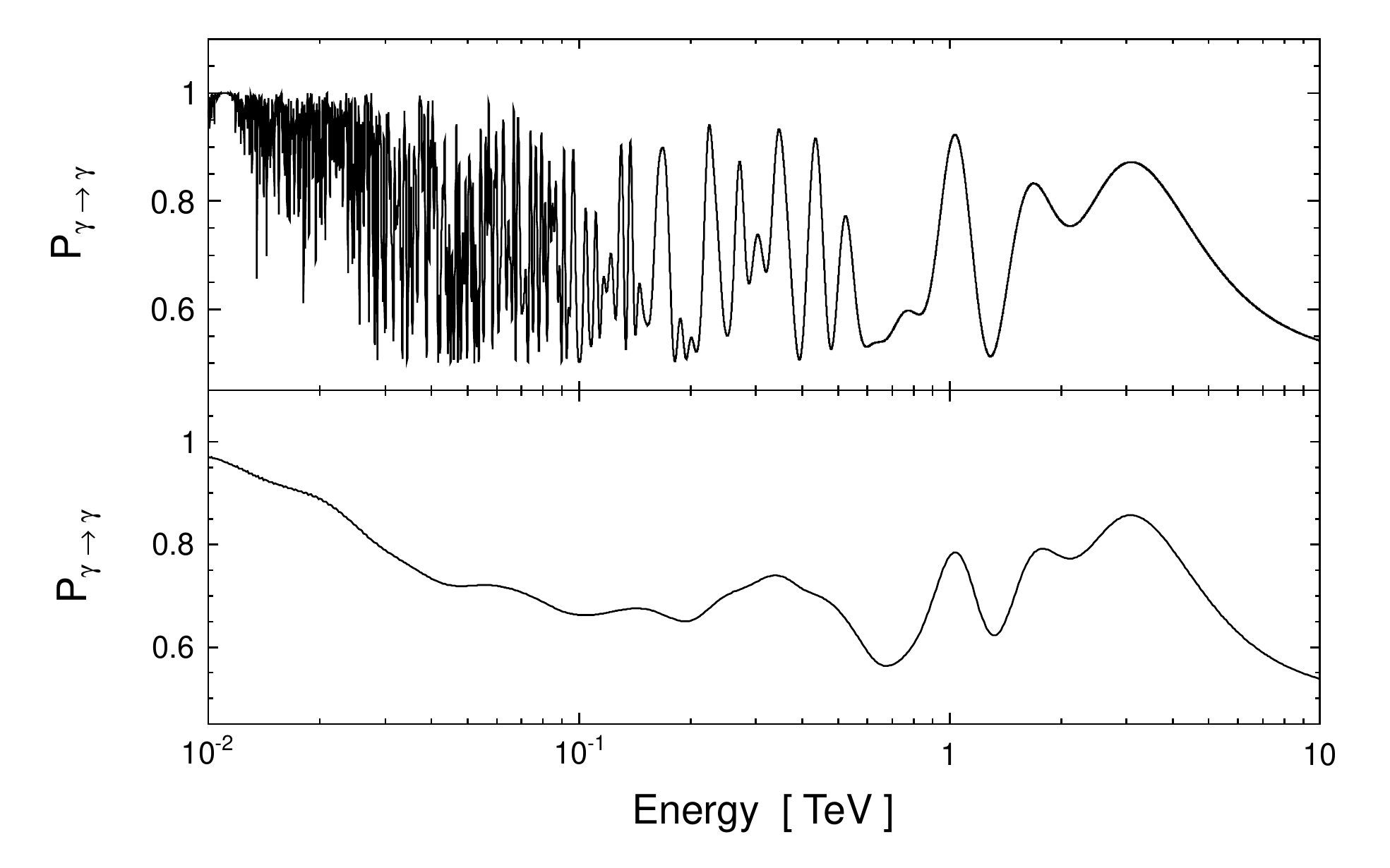}
  \caption{Transfer function for $\gamma$-rays mixing with ALPs in a galaxy cluster magnetic field (see text for details). Top panel : raw function. Bottom panel : Same function smeared with the energy resolution and bias of H.E.S.S. \label{fig:signal}}
\end{figure}

Fig.~\ref{fig:signal} shows an example of a given realization of an ALP-induced irregularity pattern in the case of an initially unpolarized photon beam (upper panel), and the same pattern smeared by the energy resolution of approximately 15 \% of H.E.S.S. (lower panel). For magnetic fields at the $\mu$G level typical of galaxy clusters and a coupling constant $g_{\gamma a} = 10^{-10}\; \rm GeV^{-1}$ close to the upper limit set by the CAST experiment~\cite{2007JCAP...04..010A}, $E_c$ is of order of 1 TeV for ALP masses of a few tens of neV. This means that the sensitivity of H.E.S.S. to ALPs lies in a limited range of mass between 10 neV and 100 neV. In the following, irregularities are searched for in the energy spectrum of the bright TeV blazar PKS 2155-304 measured by H.E.S.S~\cite{axions_hess}. 

\section{Magnetic fields on the line of sight}

PKS 2155-304 is a powerful TeV emitted BL Lac object located at redshift $z = 0.116$. This source is chosen here for two reasons. First, it has been extensively observed by H.E.S.S. and the high available statistics makes possible a precise determination of the spectrum~\cite{2005A&A...442..895A}. Second, a small galaxy cluster of radius 372 kpc is observed around this source~\cite{1993ApJ...411L..63F}, meaning that a significant magnetic field is expected in the vicinity of the source . The magnetic field in the galaxy cluster surrounding PKS 2155-304 is not measured though, leaving the possibility to make conservative assumptions about its strength and structure. Usually, Faraday rotation measurements enable to probe the structure of the magnetic field in similar galaxy clusters, hosting FR I radio galaxies (believed to be the parent population of BL Lac objects)~\cite{2002ARA&A..40..319C}. These studies report evidence for magnetic fields of the order of a few $\mu$G and turbulence in agreement with a Kolmogorov turbulence on scales as large as 10 kpc. In the following, a conservative value of $1~\mu$G is assumed for the magnetic field strength and a Kolmogorov power spectrum on scales between 1 and 10 kpc is used to describe its turbulent structure.

Conversion of photons from PKS 2155-304 in the intergalactic magnetic field (IGMF) can also be considered. The IGMF is subject to large uncertainties. The range of possible values for its strength is experimentally constrained between $10^{-16}$~G and 1 nG. In the following, a value of 1 nG is assumed for the IGMF strength to derive the corresponding ALP exclusions. This means that they are deduced from the most optimistic model. The shape of the turbulence power spectrum is not clear too. A Kolmogorov power-law may not be relevant for the description of the IGMF turbulence. A single turbulent scale of 1 Mpc is assumed to simulate the conversion in the IGMF. Conversion inside the source itself, within the jet or radio lobes is not considered here because of the very uncertain nature of the magnetic fields.

Note that each type of magnetic field adds up irregularities at TeV energies for ALPs lying in different regions of the parameter space. As we shall see in the following, constraints considering the cluster magnetic field and the IGMF can be treated independently and result in limits in different ALP mass ranges.

\begin{figure}

  \includegraphics[width=0.88\columnwidth]{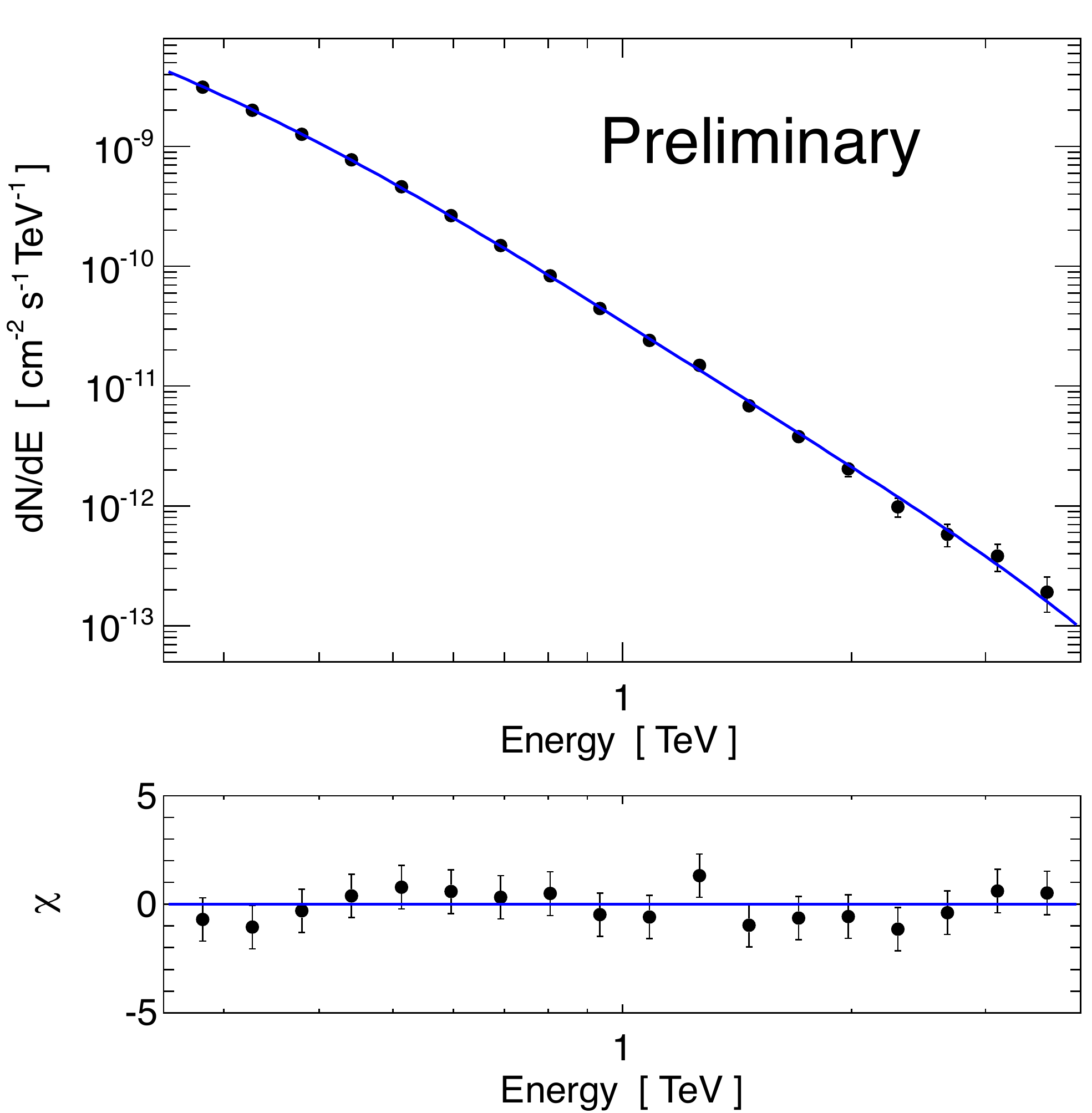}
  \centering\caption{Time-averaged energy spectrum of PKS 2155-304 (black points) for the dataset used in the analysis. Top panel: the blue line is the best fit of a curved power-law with EBL absorption to the data. Bottom panel: relative residuals of the fit normalized to the errors. \label{fig:spectrum}}
\end{figure}

\section{H.E.S.S. Observations of PKS 2155-304}

PKS 2155-304 has been observed by the H.E.S.S. phase I array of four imaging atmospheric Cherenkov telescopes that observes the $\gamma$-ray sky above a few hundreds of GeV~\cite{2006A&A...457..899A}. Observations of PKS 2155-304 with H.E.S.S. during the flare of July 2006 are selected for the spectral analysis, for a total live-time of 13 hours. Data taken during the flare of July 2006 are selected in order to minimize possible bias in the spectrum reconstruction coming from mis-subtracted background that could mimic ALP patterns.  During the high state, observations of the source are almost background free. The spectrum of the 45505 $\gamma$-ray candidates is shown on Fig.~\ref{fig:spectrum}. It is well modeled by a curved power-law convolved by absorption on the extragalactic background light (EBL) ($dN/dE \propto (E/1\rm TeV)^{-\alpha-\beta\log(E/\rm 1 TeV)}e^{-\tau_{\gamma\gamma}(E)}$) with $\alpha = 3.18 \pm 0.03_{\rm stat} \pm 0.2_{\rm syst}$, $\beta = 0.32 \pm 0.02_{\rm stat} \pm 0.05_{\rm syst}$. $\tau_{\gamma\gamma}$ is the optical depth from the EBL model of Kneiske \& Dole~\cite{2010A&A...515A..19K}. No significant irregularities appear in the spectrum so that it is used to constrain the coupling of ALPs to photons.

\section{Method and results}

\begin{figure}[h]
  \centering\includegraphics[width=0.9\columnwidth]{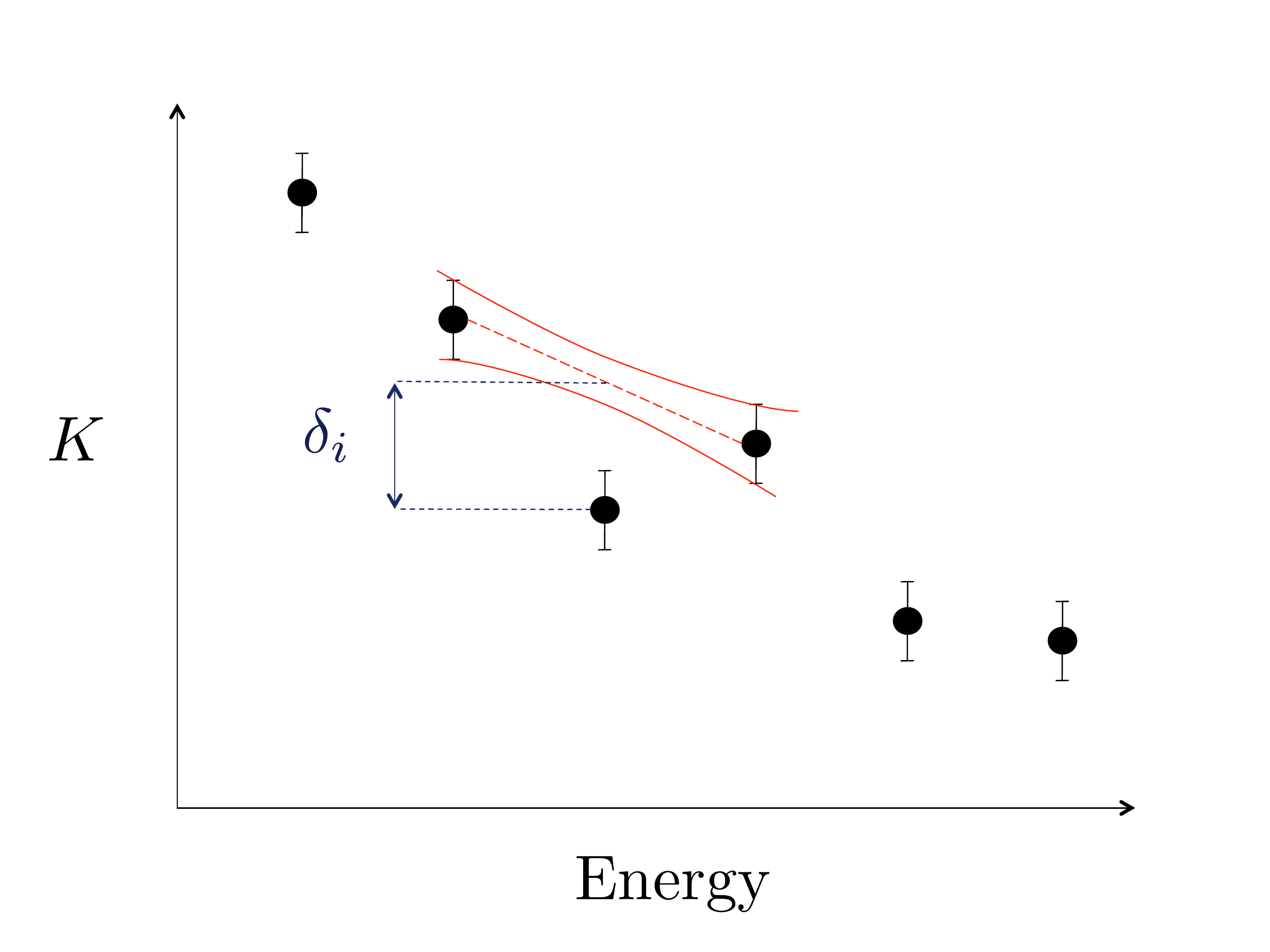}
  \caption{Schematic view of the procedure followed to estimate the level of fluctuations in the binned spectrum. \label{fig:estimator}}
\end{figure}

A possible method to constrain the value of $g_{\gamma a}$ could be to fit a spectral shape that is expected from $\gamma$-ALP oscillations on the spectrum measured by H.E.S.S. Here, the intrinsic spectrum of the source is not known, so this method would have to rely on an assumption for the spectral shape that may be wrong. In this case, the limits deduced would be biased and could be artificially better. To bypass this issue, an estimator of the irregularities in the spectrum is used that does not make the assumption of an overall spectral shape. It is based on the assumption that the spectrum should be well approximated by a power law on a scale of three consecutive bins, which is justified in the context of the astrophysical processes involved. In each group of three consecutive bins, the level of irregularity is quantified by the deviation of the middle bin from the power-law defined by the side bins as sketched on Fig.~\ref{fig:estimator}. After taking into account all errors and correlations, the deviations are quadratically summed over all the groups of three consecutive bins to form the estimator $I$ of irregularities in the spectrum. The value of $I$ may depends on the binning of the spectrum. To estimate the uncertainty on $I$, the binning is modified in size and position, which induce small variations of $I$ due to the reshuffling of some events. When applied to the previously described dataset, the average and root mean square of $I$ when varying the binning is $I = 4.10\pm0.65$. The value of 4.75 is conservatively assumed in the following to obtain the constraints.

\begin{figure}[h]
  \centering\includegraphics[width=0.9\columnwidth]{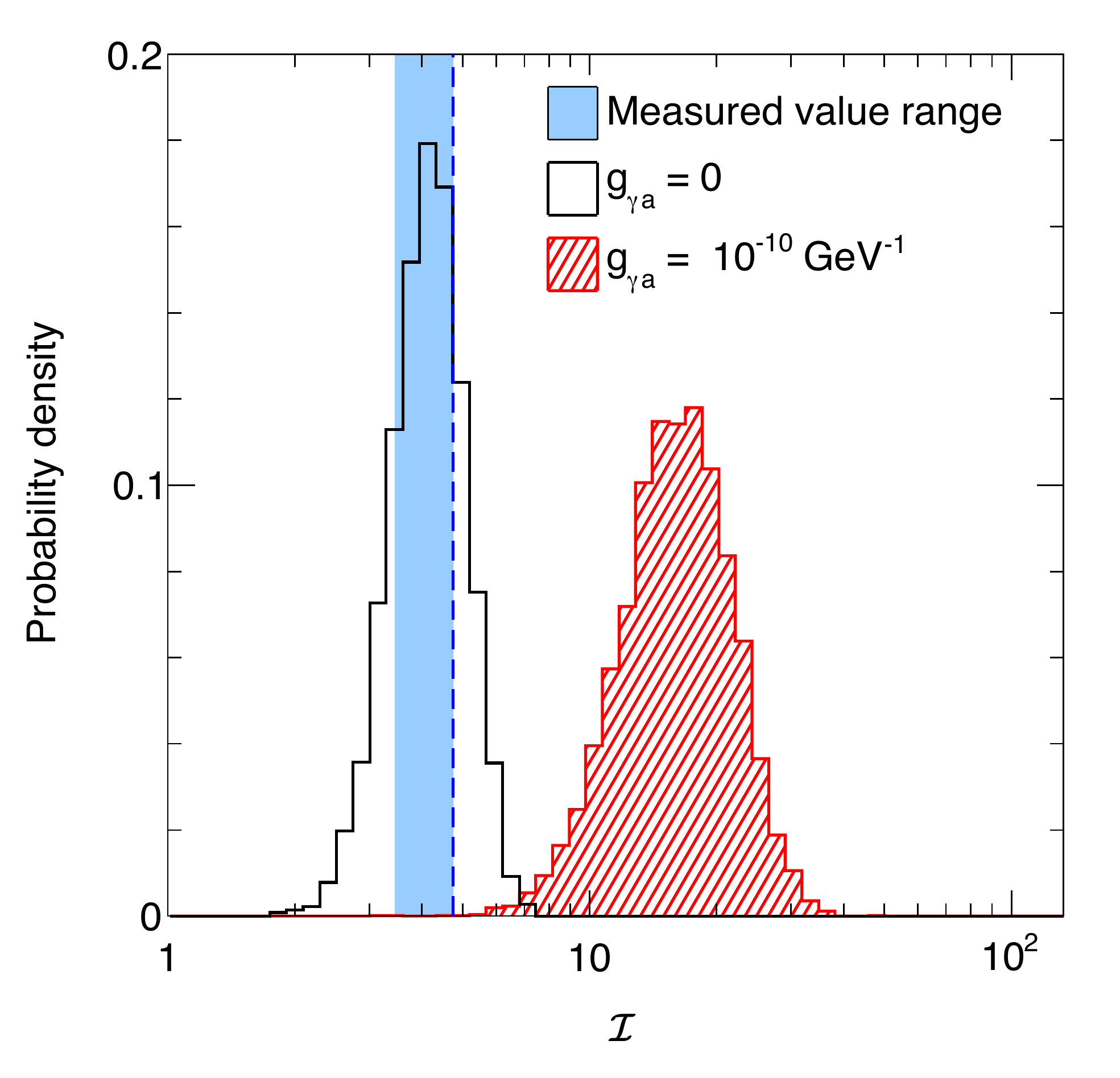}
  \caption{Predicted probability density functions of irregularity reconstructed with the fluctuation estimator for two ALP parameter sets. The vertical band correspond to measurements in the data with different bin sizes and the dashed line is the value used to set the limits. \label{fig:distrib}}
\end{figure}

In order to estimate the level of irregularities induced by ALPs that can be accommodated by the H.E.S.S. spectrum, simulations of spectra that would be observed for various ALP parameters are performed. As the exact turbulent configuration of the magnetic field is unknown, the simulations are repeated for different realizations of the magnetic field. For one parameter set, the ensemble of all the would-be measured $I$  on simulated spectra for different realizations of the magnetic field constitutes the PDF of $I$. One example of such PDFs is shown on Fig.~\ref{fig:distrib} for conversion in the galaxy cluster magnetic field and a value of the coupling strength, $g_{\gamma a} = 10^{-10}\rm GeV^{-1}$. On the same figure, the distribution for vanishing coupling is also shown. If the measured value of $I$ is excluded by the PDF at a one-sided 95\% probability, the parameters are excluded. In the case of Fig.~\ref{fig:distrib}, the value for $g_{\gamma a}$ is excluded. On that figure, the vertical blue band corresponds to the measurement of irregularities in the data with the proposed estimator. The width of the measurement band shows the amplitude of the binning related systematic error. The upper value of this interval is used to derive the limits.

\begin{figure}[h]  
  \includegraphics[width=0.9\columnwidth]{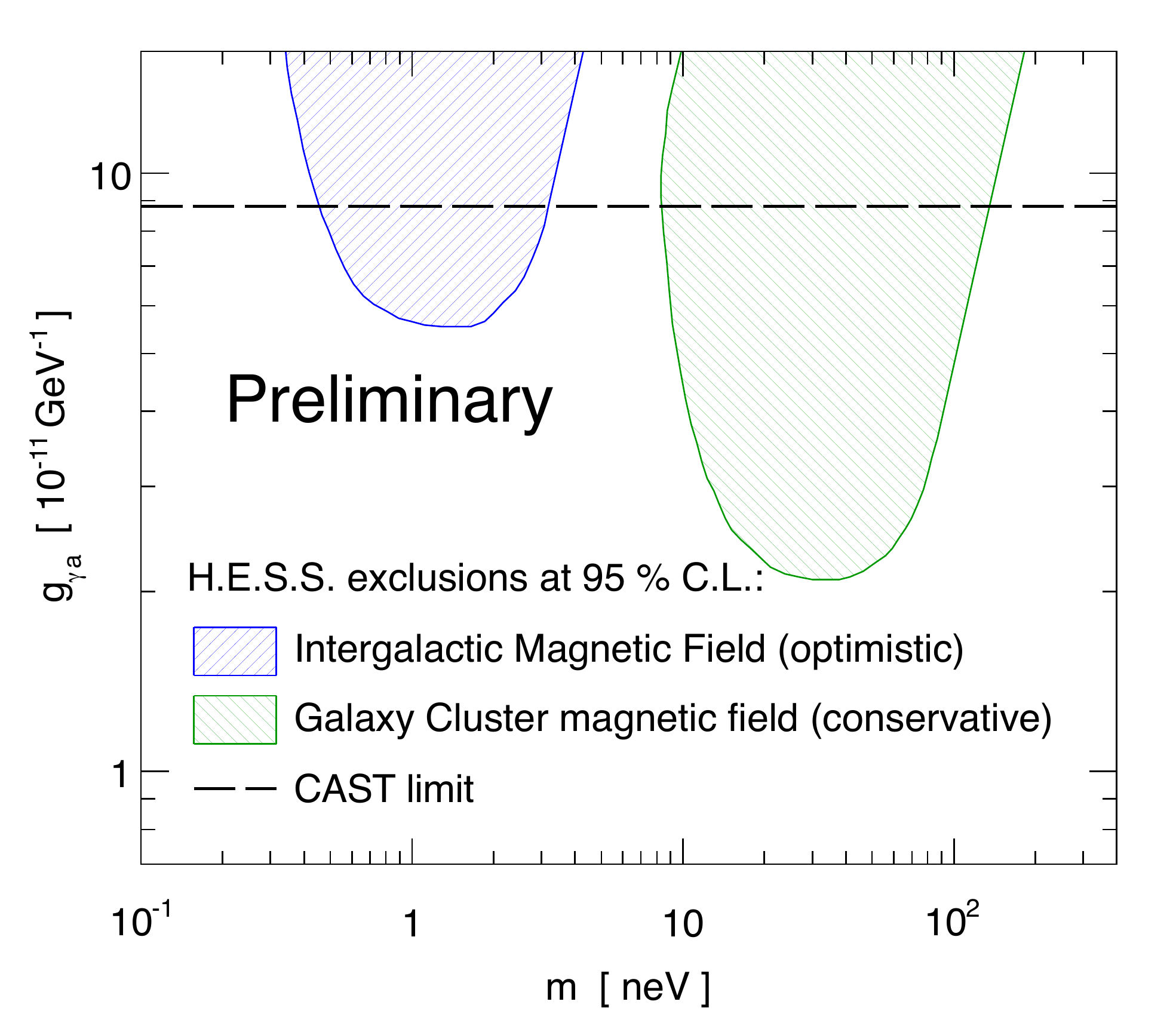}
  \centering\caption{H.E.S.S. exclusion limits on the ALP parameters $g_{\gamma a}$ and $m$. The blue dashed region on the left is obtained considering $\gamma$-ALP mixing in the IGMF with in an optimistic scenario with a 1 nG strength. The green dashed region on the right is obtained considering $\gamma$-ALP mixing in the galaxy cluster of PSK 2155-304. \label{fig:cons}}
\end{figure}


The constraints obtained with this method for conversion in the galaxy cluster and in the IGMF are shown on Fig.~\ref{fig:cons}. For the conversion in the galaxy cluster magnetic field, a conservative strength of 1 $\mu$G is assumed so that the constraints derived are considered as robust. Conversely, for the IGMF, an optimistic value of 1 nG is used and this limit is shown as an indication of the H.E.S.S. sensitivity with values that are usually assumed in the literature. The CAST limit is also shown on the plot. The H.E.S.S. exclusions improve the CAST limit in a restricted range of mass around a few tens of neV. H.E.S.S. is sensitive in a restricted mass range because the irregularities are expected in a limited energy range around the critical energy that defines the energy above which $\gamma$-ALP oscillations are possible. The range of energy that H.E.S.S. is sensitive to thus translates in a range of mass that can be probed. The limits that are shown on Fig.~\ref{fig:cons} are valid for any kind of scalar or pseudoscalar particles that couple to photons. In the future, observations including the fifth telescope of H.E.S.S. will lower the energy threshold of the spectral analysis and then enlarge the accessible mass range.

\section*{Acknowledgments}
This work benefited from the support of the French PNHE (Programme National Hautes Energies) and the French ANR program CosmoTeV.  

The support of the Namibian authorities and of the University of Namibia in facilitating the construction and operation of H.E.S.S. is gratefully acknowledged, as is the support by the German Ministry for Education and Research (BMBF), the Max Planck Society, the German Research Foundation (DFG), the French Ministry for Research, the CNRS-IN2P3 and the Astroparticle Interdisciplinary Programme of the CNRS, the U.K. Science and Technology Facilities Council (STFC), the IPNP of the Charles University, the Czech Science Foundation, the Polish Ministry of Science and  Higher Education, the South African zDepartment of Science and Technology and National Research Foundation, and by the University of Namibia. We appreciate the excellent work of the technical support staff in Berlin, Durham, Hamburg, Heidelberg, Palaiseau, Paris, Saclay, and in Namibia in the construction and operation of the equipment.

\end{document}